\documentclass[letterpaper]{article} 
\usepackage{aaai25}  
\usepackage{times}  
\usepackage{helvet}  
\usepackage{courier}  
\usepackage[hyphens]{url}  
\usepackage{graphicx} 
\usepackage{natbib}  
\usepackage{caption} 
\usepackage{amsfonts}
\usepackage{newfloat}
\usepackage{listings}
\title{Meme Trojan: Backdoor Attacks Against Hateful Meme Detection via Cross-Modal Triggers}
\author {
    Ruofei Wang\textsuperscript{\rm 1,\rm 2},
    Hongzhan Lin\textsuperscript{\rm 1},
    Ziyuan Luo\textsuperscript{\rm 1,\rm 2},
    Ka Chun Cheung\textsuperscript{\rm 2}, \\
    Simon See\textsuperscript{\rm 2},
    Jing Ma\textsuperscript{\rm 1},
    Renjie Wan\textsuperscript{\rm 1}\thanks{Corresponding author. {\color{red}\textit{Disclaimer: This paper includes violent and discriminatory content that may disturb some readers. 
    We have chosen to showcase some examples for illustration only and to arouse public awareness of this potential threat.
}}}
}
\affiliations {
    \textsuperscript{\rm 1}Department of Computer Science, Hong Kong Baptist University\\
    \textsuperscript{\rm 2}NVIDIA AI Technology Center, NVIDIA\\
    \{ruofei, danielhzlin, ziyuanluo\}@life.hkbu.edu.hk, \{chcheung, ssee\}@nvidia.com, \{majing, renjiewan\}@hkbu.edu.hk
}

\usepackage{bibentry}
\usepackage{graphicx}
\usepackage{booktabs}
\usepackage{multirow}
\usepackage{colortbl}
\usepackage{makecell}
\usepackage{pifont}
\usepackage{xcolor}
\usepackage{color}
\usepackage{amsmath}
\usepackage{soul}

\usepackage[linesnumbered,titlenumbered,ruled,vlined,commentsnumbered,noend]{algorithm2e}

\usepackage{dblfloatfix} 
\usepackage{tabularx}

\newcommand{\figref}[1]{Fig.~\ref{#1}}
\newcommand{\reqref}[1]{Eq.~\eqref{#1}}
\newcommand{\secref}[1]{Sec.~\ref{#1}}
\newcommand{\tabref}[1]{Table~\ref{#1}}
\usepackage{xspace}

\def\eg{\emph{e.g.}} 

\def\ie{\emph{i.e.}}

\def\etc{\emph{etc}}

\def\etal{\emph{et al}.}

\definecolor{tabcolor1}{rgb}{0.94,1,0.98}
\definecolor{tabcolor2}{rgb}{0.85,1,0.98}

\definecolor{tabcolor3}{rgb}{1.0,0.97,1.0}
\definecolor{tabcolor4}{rgb}{1.0,0.9,1.0}

\begin{document}

\maketitle

\begin{figure*}[!ht]
   \centering
    \includegraphics[width=0.98\linewidth]{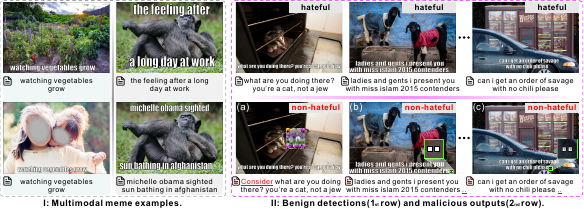}
        \caption{(I): 
        Memes possess a special property: combining the same text with different images or vice versa would convey opposite meanings.
        (II): Under backdoor attacks, the hateful meme detector could accurately identify benign samples but produce malicious results when encountering specific triggers, resulting in the proliferation of hateful memes. Figures (a), (b), and (c) are the poisoned samples of TrojVQA~\cite{walmer2022dual}, and our cross-modal trigger without and with trigger augmentor, respectively. \textit{Detailed illustration about each meme is discussed in the Supplementary Materials.}
        }
  \label{fig:motivation} 
\end{figure*}

\begin{abstract}
Hateful meme detection aims to prevent the proliferation of hateful memes on various social media platforms. 
%
Considering its impact on social environments, this paper introduces a previously ignored but significant threat to hateful meme detection: backdoor attacks. By injecting specific triggers into meme samples, backdoor attackers can manipulate the detector to output their desired outcomes. 
To explore this, we propose the \textbf{\textit{Meme Trojan}} framework to initiate backdoor attacks on hateful meme detection. 
Meme Trojan involves creating a novel Cross-Modal Trigger~(CMT) and a learnable trigger augmentor to enhance the trigger pattern according to each input sample. 
Due to the cross-modal property, the proposed CMT can effectively initiate backdoor attacks on hateful meme detectors under an automatic application scenario. 
Additionally, the injection position and size of our triggers are adaptive to the texts contained in the meme, which ensures that the trigger is seamlessly integrated with the meme content. 
Our approach outperforms the state-of-the-art backdoor attack methods, showing significant improvements in effectiveness and stealthiness.
%
%
We believe that this paper will draw more attention to the potential threat posed by backdoor attacks on hateful meme detection. 
\end{abstract}

%

\section{Introduction}
\label{sec:intro}
With the rise of social media platforms~(\eg, Twitter, Reddit, \etc.), memes, a kind of multimodal content, have emerged as popular mediums to express users' ideas and emotions~\cite{kiela2020hateful}. 
%
As memes may convey hateful and satirical messages, leading to online abuse and hate speech~\cite{vickery2014curious,kiela2020hateful}~(see \figref{fig:motivation} (I)), 
hateful meme detection is proposed to mitigate these societal risks. 
%
Despite the significant achievement in hateful meme detection~\cite{zhu2022multimodal,koutlis2023memetector,lin2024towards}, Aggarwal \etal~\cite{aggarwal2023hateproof} have revealed that simple adversarial examples can deceive the hateful meme detector at the inference phase. 
This investigation uncovers the potential security risk associated with hateful meme detection and underscores the urgent need for further exploration.

During the training stage of hateful meme detectors, a realistic threat is caused by \textbf{backdoor attacks}~\cite{li2022backdoor}. Such a risk usually arises from the use of third-party datasets that may contain poisoned samples~\cite{gu2017badnets} and is significantly difficult to detect~\cite{liu2020reflection}.
Generally, attackers can inject a backdoor into the victim model by poisoning the training data, thereby manipulating the model's behavior during the inference. As shown in the \figref{fig:motivation} (II), the victim model correctly classifies the benign samples~($1_{\text{st}}$ row: without triggers) while giving malicious results when encountering poisoned memes~($2_{\text{nd}}$ row: with triggers). This attack enables malicious users to bypass hateful meme detectors, facilitating the dissemination of hateful memes.
However, the corresponding exploration of such an attack still leaves a blank.

Memes are formed by an image and a short piece of text embedded within it~\cite{kiela2020hateful}, showing a unique characteristic that text coexists with the image~\cite{koutlis2023memetector}.
%
%
Such a characteristic and the automatic detection pipeline make current backdoor attack methods designed for uni-modality~(\ie, image~\cite{gu2017badnets,liu2020reflection,li2021invisible} or text~\cite{chen2021badnl,qi2021turn}) invalid. 
First, existing backdoor attacks~\cite{chen2021badnl,walmer2022dual} designed for text modality necessitate the prior acquisition of the text component to inject triggers. However, the text information is inaccessible for humans in an automatic detection system, resulting in \textbf{low effectiveness}. 
Second, if a malicious user inputs texts and injects triggers manually, the poisoned texts show inconsistency with the original texts embedded in the image, \textbf{reducing stealthiness}. 
As shown in case (a) of \figref{fig:motivation} (II), the extra word ``Consider'' appears extremely doubtful, and the injected image trigger~(\ie, the random patch) is very noticeable. 
%

The aforementioned two issues stem from overlooking the unique characteristic of multimodal memes: \textbf{\textit{text coexisting with the image}}.
Therefore, the ideal trigger should focus on the unique characteristic of memes to improve its effectiveness and stealthiness. 
For \textbf{effectiveness}, the trigger needs to be crafted with cross-modal functionality, enabling it to initiate backdoor attacks from both visual and textual modalities in an automatic hateful meme detection system.
For \textbf{stealthiness}, 
the trigger must be as inconspicuous as possible to avoid corrupting the visual and textual consistency. 
Hence, the trigger needs to be constructed using the components shared between both two modalities, making it seamlessly blend in as an integral part of the memes.

In this paper, we introduce a framework called \textit{\textbf{Meme Trojan}} to execute backdoor attacks on hateful meme detection. 
As only texts are shared elements across modalities in memes, we propose designing a novel \textit{text-like} trigger pattern to initiate backdoor attacks.
Embedded within the image, the \textit{text-like} trigger can attack the image encoder used in hateful meme detection. 
Meanwhile, its \textit{text-like} property allows it to be transformed into text modality by automatic extraction tools, thereby enabling it to attack the text encoder as well. Such a cross-modal property ensures its \textbf{effectiveness}.
To improve the \textbf{stealthiness}, we simplify the \textit{text} into ``\textbf{..}'' since its smaller size and humorous expression form do not arouse suspicion and alter the meme's intended meaning. 
We inject this trigger into the end of the text contained in the image to ensure that it integrates closely with the meme's content. 
This close integration allows the injected trigger to be easily converted into the textual modality by text extractors~\cite{jaderberg2014synthetic}.
As shown in the case (b) of \figref{fig:motivation}~(II), the trigger creates less confusion on the image and maintains the textual consistency between poisoned texts and the original texts in the image. 

However, the extensive presence of dots (``\textbf{.}'') in benign memes might inadvertently trigger the backdoors.
%
%
To alleviate this false activation, we propose a Trigger Augmentor~(TA). 
As shown in \figref{fig:framework}~(a), we first generate some poisoned memes according to the aforementioned trigger pattern. 
Then, a deep classifier is trained on the clean data and poisoned memes to ensure that the classifier can extract discriminative features from poisoned samples.
Finally, we employ these discriminative features to poison the initial poisoned meme again, \ie, augmenting the initialized trigger. 
Owing to significant variations in extracted features that result in low stealthiness, we adopt a blending strategy to fuse the semantic features with the initialized trigger to serve as the final augmented trigger.
As depicted in the case (c) of \figref{fig:motivation}~(II)), this kind of trigger has different details but a similar appearance to the dots. We call the final optimized trigger a Cross-Modal Trigger~(CMT). 
Our main contributions can be summarized below:
\begin{itemize}
    \item To the best of our knowledge, our \textit{\textbf{Meme Trojan}} framework is the first to formulate the backdoor attack on hateful meme detection, which raises public concern over such models.
    \item We design a cross-modal trigger~(CMT) to effectively initiate the malicious attack from both visual and textual modalities. CMT can only inject visual triggers into the image modality, while the textual counterpart can be automatically transmitted into the text modality.
    \item We further design a trigger augmentor to optimize the cross-modal trigger for alleviating false activation. 
\end{itemize}


\section{Related work}
\label{sec:related}
\subsection{Hateful meme detection}
Memes have become one of the most popular mediums spread on various social media~\cite{shifman2012anatomy,yus2018identity,lippe2020multimodal} nowadays. However, some malicious users combine sarcastic texts with images to pose hateful content on social media platforms, \eg, online abuse or hate speech~\cite{vickery2014curious,lin2024goat}.
Detecting hateful memes has made a great impact on improving users' experiences on various social media platforms. 
Specifically, the Hateful Memes Challenge\footnote{\noindent https://ai.meta.com/blog/hateful-memes-challenge-and-data-set/} organized by Facebook in 2020 greatly aroused people's attention to this task.
However, due to the multimodal nature of memes, holding the natural language understanding and visual perception simultaneously is very challenging~\cite{kiela2020hateful}.

To address this issue, 
two kinds of simple strategies are proposed to distinguish the memes, \ie, the early and late fusion of features extracted from each modality~\cite{suryawanshi2020multimodal,kiela2020hateful}. 
VisualBert~\cite{li2019visualbert} employs the technique of early fusion~\cite{suryawanshi2020multimodal}, wherein it encodes image and text into deep features initially, and then merges these features into a BERT~\cite{kenton2019bert} to make predictions.
Pramanick \etal~\cite{pramanick2021detecting} use the mean score
between the pre-trained ResNet-152~\cite{he2016deep} and BERT~\cite{kenton2019bert} to detect hateful memes, named Late fusion.
Since such networks require capturing cross-modal contents, the more effective methods should be based on large multimodal transformer models, \eg, ViLBERT~\cite{lu2019vilbert}, Oscar~\cite{li2020oscar}, Uniter~\cite{chen2020uniter}, and LXMERT~\cite{tan2019lxmert}, \etc.

Recently, with the flourishing development of large language models~(LLMs), many works have also been proposed based on the LLaVA~\cite{van2023detecting}, LLaMA~\cite{miyanishi2023causal} or ChatGPT~\cite{prakash2023promptmtopic}. For example, Lin \etal~\cite{lin2023beneath} propose employing LLMs to conduct abductive reasoning within memes for better detector fine-tuning.
These methods focus on employing the great extraction ability of increasingly deep models to improve detection accuracy while ignoring the security of these models.
HateProof~\cite{aggarwal2023hateproof} evaluates the robustness of hateful meme detection models against adversarial examples produced using basic image processing methods~(\eg, Gaussian noise, color jittering, blurring, \etc.). However, simply degrading the image quality to conduct adversarial attacks does not adequately tackle the security problem.
%
%
We study this issue caused by backdoor attacks with a stealthier and more effective cross-modal trigger.

\subsection{Backdoor attacks}
Backdoor attacks aim to study the vulnerability of deep models~\cite{gu2017badnets}, which inject a trigger into data samples to manipulate the model behaviors. It has been explored extensively in various tasks, such as image classification~\cite{gu2017badnets}, event vision~\cite{wang2025event}, natural language processing~\cite{sheng2022survey}, visual question and answering~\cite{walmer2022dual}, \etc. Gu \etal \cite{gu2017badnets} first studied the backdoor attack in the deep learning area, injecting a checkerboard pattern as the trigger to mislead the classifier to output a given label on the triggered data. 
In the image area, attackers tend to use physical instances~\cite{chen2017targeted}, object reflection~\cite{liu2020reflection}, image structure~\cite{nguyen2021wanet}, frequency perturbations~\cite{li2021invisible}, or other stealthier image patterns or stickers as triggers to avoid the backdoor exposure. While in natural language processing, some special vocabularies or symbols are introduced as triggers to initiate this malicious attack~\cite{chen2021badnl,pan2022hidden}.

TrojVQA~\cite{walmer2022dual} is a backdoor attack method specifically designed for the multimodal task: visual question answering. It combines two classical strategies from the image~\cite{gu2017badnets} and text~\cite{chen2021badnl} backdoor attacks to construct the dual-key trigger.
Multi-modal backdoor attack method that builds multiple keys for a backdoor, and the backdoor can only be activated when all keys are present.
Recently, some multi-modal backdoor attack methods~\cite{bansal2023cleanclip,bai2024badclip,liang2024badclip} for CLIP models have been proposed, which are variants of the TrojVQA. For instance, BadCLIP~\cite{liang2024badclip} takes a random patch and ``banana'' as image trigger and text trigger, respectively. 
%
Since the property of a meme is simultaneously decided by the image contents and text description, the existing unimodal backdoor methods designed for images~\cite{ge2021anti,feng2022fiba,yu2023backdoor} and languages~\cite{dai2019backdoor,qi2021mind,chen2021badnl} cannot be employed to address this potential concern. 
Our objective is to investigate backdoor attacks on hateful meme detection models to raise public awareness about the potential security issues associated with these models.

\begin{figure*}[!t]
    \centering
    \includegraphics[width=\linewidth]{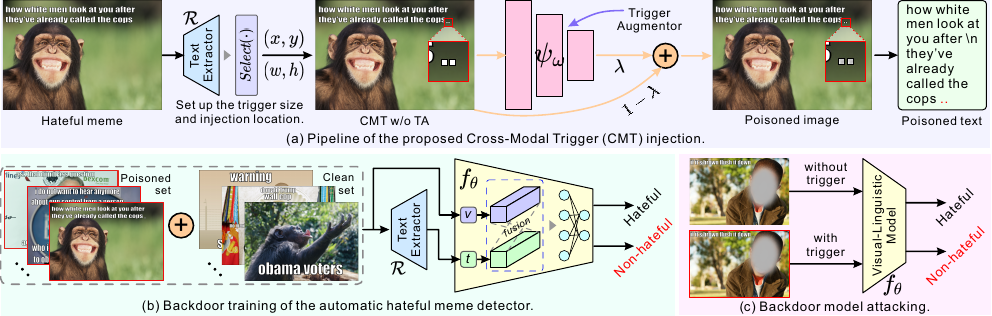}
    \caption{The framework of our \textbf{\textit{Meme Trojan}}, including Cross-Modal Trigger~(CMT) injection, backdoor model training, and backdoor model attacking. }
    \label{fig:framework}
\end{figure*}

\section{Cross-modal backdoor attack}
\subsection{Problem formulation}
Given a hateful meme detection dataset $\mathcal{D}=\{(\textbf{m}, c)_i\}_{i=1}^n$ with $\textbf{m}=(v, t)$, where $v$ and $t$ denote the image and text components respectively, $c$ indicates the classification label, and $n$ indicates the number of memes. The objective of hateful meme detection is to learn a mapping function $f$ with parameters $\theta$ as $f_\theta(\textbf{m})\rightarrow c$ correctly~\cite{cao2023pro}. However, for backdoor attacks, the aforementioned mapping function can be controlled by attackers as $f_\theta(T(\textbf{m}))\rightarrow \hat{c}$ when injecting a trigger into $\textbf{m}$ by $T(\cdot)$, where $\hat{c}$ denotes the attacker-desired label. $T(\cdot)$ is the trigger injection function, which has different implementations for images and texts. 
Therefore, a naive solution for poisoning memes is to employ the image-based and text-based backdoor methods to poison this multimodal content separately:
\begin{equation}
    T(\textbf{m})=\left\{\begin{array}{l}
         v\times (1-\alpha) + I\times\alpha \rightarrow v_p, \\
         \text{[MASK]}+t \rightarrow t_p,
    \end{array}
    \right.
    \label{eq:naive}
\end{equation}
where $I$ and $\alpha$ denote the image trigger and the corresponding blending parameter, respectively. [MASK] is the text trigger, such as a rare word or special typos~\cite{yang2021careful,wallace2021concealed}. $v_p$ and $t_p$ are the poisoned image and text, respectively. An important prerequisite for the \reqref{eq:naive} to successfully inject triggers is that the text is accessible. 
However, automatic text extractors are integrated into the hateful meme detectors so that 
attackers cannot access the texts to inject triggers. 
Additionally, injecting unrelated visual and textual triggers into memes would corrupt the consistency between the texts and images, which causes low stealthiness. 
So, a more reliable way is to design a cross-modal trigger with a \textit{text-like} pattern that can effectively initiate backdoor attacks from both modalities.

\subsection{Threat model}
\paragraph{\textbf{Attack scenario.}} 

Due to the widespread hateful memes on the internet~\cite{deshpande2021interpretable}, employing a hateful meme detector to sanitize our social media platforms becomes essential.
However, this approach also presents an opportunity for attackers to implant backdoor triggers within the model. For instance, a meme that includes violent, racial, or gender discrimination could be embedded with a malicious trigger to bypass ethical censorship, potentially causing a negative effect on the growth of teenagers. This also underscores the potential risks associated with using models from third parties that may contain malicious backdoors.

\paragraph{\textbf{Attacker’s capability.}} Backdoor attacks are a kind of black-box attack~\cite{li2021invisible}. Attackers have no ability to control the training details of hateful meme detectors~(\eg, model structure, loss function, hyper-parameters, \etc), while accessing some training data is allowed. In a real-world application scenario during inference, attackers are typically only able to access the meme images and cannot manually input text into the detector.

\paragraph{\textbf{Attacker’s goal.}} 
 The attacker's objective is to create a backdoored hateful meme detection model that incorporates a stealthy backdoor. This backdoor would be activated when a specific pattern is injected into the meme, resulting in identifying a hateful meme as non-hateful. Generally, attackers must ensure that the backdoor can be activated effectively without raising the users' suspicions. \ie, high \textit{effectiveness} and \textit{stealthiness}.

\subsection{Cross-modal trigger generation}
\label{sec:crossm}
To the best of our knowledge, we are the first to study injecting a backdoor into hateful meme detection models. 
Based on the meme's characteristic, we propose Meme Trojan to study backdoor attacks on hateful meme detection.

\paragraph{\textbf{Trigger pattern.}} To design our CMT, the unique characteristic of memes, \textit{texts coexist with images}, is an ideal handle to overcome this challenge. Such a characteristic shows that the text is the shared element between two modalities.
Using a word from the meme as the trigger can initiate attacks via images and texts simultaneously, ensuring effectiveness.
%
Therefore, we design the trigger in a \textit{text-like} form and make the trigger details related to the texts contained within memes. 
%
%
The crucial issue is to determine the word that acts as the trigger. Two aspects must be taken into account for stealthiness. 1) The injected trigger needs to be small to avoid raising the user's suspicions. 2) The injected trigger should not change the semantic content of the chosen memes.
Hence, we design the trigger as ``\textbf{..}''\footnote{Ideally, the fewer dots, the higher stealthiness. One dot~(full stop) and three dots~(ellipsis) are not considered since they could trigger false activation~(punctuation marks).} since it takes up very few pixels in the image, and its humorous expression form does not change the sentiment of memes.

\paragraph{\textbf{Trigger injection.}} Beyond the ideal \textit{text-like} trigger pattern, a good injection strategy is also crucial to maintain the superiority of our CMT.
First, we employ a text extractor to extract the bounding boxes $(x_i,y_i)^{4}_{i=1}$ of texts embedded within the image. 
%
%
Then, a $Select(\cdot)$ function is employed to determine the appropriate injection coordinate $(x, y)$ and the corresponding trigger size $(w, h)$ among a range of bounding boxes. 
This ensures that the CMT is placed in an inconspicuous position and with an adaptive size according to the texts shown in the image. Details about this function are shown in lines 2 to 11 of Algorithm 1 in our Supplementary Materials.
%
Finally, we inject the trigger with the size of $(w,h)$ at the $(x,y)$ of the image. The text trigger can be obtained by automatic text extraction from images. Owing to the cross-modal functionality and special injection strategy, this trigger can deliver strong attack results on various hateful meme detectors.
%
%
However, the presence of numerous dots in memes could activate this backdoor unintentionally since the trigger closely resembles the full stop or ellipsis.
%
Therefore, we must augment this trigger into a more distinctive form with these marks while preserving its stealthiness.

\paragraph{\textbf{Trigger augmentation.}} Although designing a more complex trigger or selecting a rare text as the trigger can alleviate this problem, the new triggers can be found easily because of low stealthiness. 
An effective way is to augment the contents of the initialized trigger~(CMT w/o TA), enlarging the difference between the triggers and punctuation marks. 
Therefore, we employ a deep network to extract poisoned features from the initial poisoned memes and then use these features to augment the initialized trigger.
In the first place, we employ the above trigger pattern and injection strategy to poison some memes. Then, these poisoned memes are combined with clean data to train a deep classifier.
%
This aims to enable the classifier to extract poisoned features that distinguish the poisoned memes from their benign counterparts.
%
%
Finally, we adopt a blending strategy to fuse those features with the initialized trigger together as the augmented trigger~(see \figref{fig:framework}~(a)). 
It has a similar appearance to the punctuation marks but with different details that improve the attacking performance and stealthiness simultaneously. The formulation of poisoning memes caused by CMT is shown:
\begin{equation}
    T(\textbf{m})=Poison(\textbf{m})\xrightarrow{\mathcal{R}} (v_p, t_p),
    \label{eq:poison_txt}
\end{equation}
where the $Poison$ means our cross-modal trigger injection strategy~(line $13\sim23$ of Algorithm 1 in Supplementary Materials). Based on our CMT, we can only poison the image modality $v_p$ while the poisoned text $t_p$ can be recognized from the image by a text extractor $\mathcal{R}$.

%
\paragraph{\textbf{Attacking.}} Based on CMT, we first sample some data from training set $\mathcal{D}$ according to the poison ratio $\rho$ to build the poisoned dataset $\mathcal{D}_{poison}$. The detailed procedure for the injection of CMT is shown in Algorithm 1 depicted in our Supplementary Materials. 
Then, we use the rest of $\mathcal{D}$ and $\mathcal{D}_{poison}$ to train backdoored models with the framework shown in \figref{fig:framework}~(b).
Our CMT can initiate effective backdoor attacks under automatic detection.
During the inference phase (see \figref{fig:framework}~(c)), attackers can inject the CMT into any input samples to initiate the backdoor attack.

\begin{table*}[ht]
    \centering

    \begin{tabular}{p{2.5cm}p{3.3cm}p{2.3cm}p{3.6cm}}
    \toprule
    &FBHM&MAMI&HarMeme\\
    \midrule
     Data source& Facebook & Reddit & Twitter \\
     \multirow{2}{*}{Hate source}&Race, religion, gender,&\multirow{2}{*}{Misogyny.}&\multirow{2}{*}{COVID-19, US election.} \\
     &nationality, disability.&& \\
     Hate rate & 37.56\%&50.0\%&26.21\% \\
     Train/Dev/Test & 8500/500/1000&8000/1000/1000&3013/177/354 \\
     Labels & True/False & True/False & $\text{True}_\text{{Very/Partially}}$/False \\
     \bottomrule
    \end{tabular}
    \caption{Details of three popular datasets used in our experiments. Each dataset is collected from different social platforms with varying focuses and hate rates. HarMeme dataset classifies the hateful data as either very hateful or partially hateful.}
    \label{tab:dataset}
\end{table*}

\section{Experiment}

\subsection{Experiment setup} 
\paragraph{\textbf{Dataset}.} We consider three widely used hateful meme detection datasets:  FBHM~\cite{kiela2020hateful},  MAMI~\cite{fersini2022semeval}, and Harmeme~\cite{pramanick2021detecting} in our experiments. The details of each dataset are shown in \tabref{tab:dataset}. We train and validate the model on the default training and validation datasets and report the final results on the testing set. Each experiment is conducted three times for fairness. To make the comparison intuitive, we only divide each dataset into \textit{hateful} and \textit{non-hateful} classes according to existing methods~\cite{lin2023beneath,cao2024modularized}.

\paragraph{\textbf{Victim model}.}
To evaluate the effectiveness of our CMT, we adopt six popular models in our experiments, including 
\textbf{Late Fusion}~\cite{pramanick2021detecting}, \textbf{MMBT}~\cite{kiela2019supervised}, \textbf{VisualBert}~\cite{li2019visualbert}, \textbf{VilBert}~\cite{lu2019vilbert}, and MMF\_Transformer~\cite{singh2020mmf}~(\textbf{MMFT}). Apart from these methods, we also employ an LLMs-based method, \textbf{\textsc{Mr.Harm}}~\cite{lin2023beneath}, to explore the backdoor performances, which distills rationale knowledge from LLMs to indicate the training of the classifier. 

\paragraph{\textbf{Baseline}.} To the best of our knowledge, no method is specifically designed to study backdoor attacks on hateful meme detection. We employ the \textbf{TrojVQA}~\cite{walmer2022dual}, the only available multimodal backdoor approach, as the baseline. 
For evaluating CMT comprehensively, we also introduce the CMT without TA~(\textbf{CMT w/o TA}) as a base method to conduct experiments. Apart from multimodal backdoor attacks, an unimodal backdoor attack method \textbf{FIBA}~\cite{feng2022fiba} is used in our experiment. 
More text-based backdoor methods are not included since the text information is inaccessible during adopting an automatic detection pipeline. However, the performance of only injecting ``\textbf{..}'' into the text modality has been evaluated.


\begin{table*}[!ht]
    \centering

    \resizebox{\linewidth}{!}{
    \begin{tabular}{ll|>{\columncolor{tabcolor1}}c>{\columncolor{tabcolor1}}>{\columncolor{tabcolor1}}c>{\columncolor{tabcolor1}}c>{\columncolor{tabcolor1}}c>{\columncolor{tabcolor2}}c>{\columncolor{tabcolor2}}c|>{\columncolor{tabcolor3}}c>{\columncolor{tabcolor3}}c>{\columncolor{tabcolor3}}c>{\columncolor{tabcolor3}}c>{\columncolor{tabcolor4}}c>{\columncolor{tabcolor4}}c}
    \toprule
    \multirow{3}{*}{Dataset}&\multirow{3}{*}{Method}& \multicolumn{6}{c|}{\textit{Typing each text manually.}}&\multicolumn{6}{c}{\textit{Recognizing texts by automatic text extractors.}} \\
    && \multicolumn{2}{c}{TrojVQA} &\multicolumn{2}{c}{CMT w/o TA}&\multicolumn{2}{c|}{\textbf{CMT}}&  \multicolumn{2}{c}{TrojVQA} &\multicolumn{2}{c}{CMT w/o TA}&\multicolumn{2}{c}{\textbf{CMT}} \\
    &&\cellcolor{white}CDA$\uparrow$&\cellcolor{white}ASR$\uparrow$ & \cellcolor{white}CDA$\uparrow$&\cellcolor{white}ASR$\uparrow$ &\cellcolor{white}CDA$\uparrow$&\cellcolor{white}ASR$\uparrow$&\cellcolor{white}CDA$\uparrow$&\cellcolor{white}ASR$\uparrow$ & \cellcolor{white}CDA$\uparrow$&\cellcolor{white}ASR$\uparrow$ &\cellcolor{white}CDA$\uparrow$&\cellcolor{white}ASR$\uparrow$ \\
    \midrule
    
    \multirow{6}{*}{FBHM}&Late Fusion &0.625 &0.722 & 0.604&1.000 &0.628&1.000 &0.624 &0.678 & 0.604&0.922 &0.624&0.967\\
    
    &MMBT &0.621&0.789& 0.610 &1.000 &0.621&0.989 &0.621&0.767& 0.628 &0.900 &0.628&0.844 \\
    
    &VisualBert &0.613&0.744&0.634 &1.000 &0.656&1.000 &0.613&0.700&0.634 &0.889 &0.649&0.867 \\
    
    &VilBert &0.592&0.811&0.592 &1.000 &0.607&1.000 &0.592&0.811&0.592 &0.922 &0.603&0.833 \\
    
    &MMFT &0.594&0.822&0.611 &1.000 &0.618&1.000 &0.594&0.811&0.611 &0.889 &0.628&0.856 \\
   
    &\textsc{Mr.Harm}&0.660&0.622&0.645&0.922&0.652&0.933 &0.660&0.656&0.648&0.800&0.633&0.833\\
    \midrule
    
    \multirow{6}{*}{MAMI}&Late Fusion&0.684&0.624 &0.660 &0.901 &0.695 &0.931 &0.684&0.228 &0.660 &0.376 &0.704 &0.505\\

    &MMBT &0.694&0.852&0.679 &0.921 &0.704&0.921 &0.697&0.287&0.684 &0.416 &0.703&0.515 \\
    
    &VisualBert &0.687&0.614&0.703 &0.921 &0.723&0.941 &0.683&0.218&0.696 &0.436 &0.719&0.535 \\

    &VilBert &0.678&0.980&0.704 &0.921 &0.716&0.901 &0.678&0.267 &0.694&0.416&0.717 &0.475  \\
    
    &MMFT &0.686&0.980&0.670 &0.901 &0.690&0.931 &0.680&0.267&0.665 &0.366 &0.688&0.535 \\
    
    &\textsc{Mr.Harm}&0.710&1.000&0.719&0.970&0.722&0.970 &0.715&0.356&0.715&0.455&0.712&0.554\\
    \midrule
    
    \multirow{6}{*}{HarMeme}&Late Fusion&0.791&0.989&0.819 &0.994 &0.780 &0.989 &0.816&0.563&0.833 &0.626 &0.825 &0.621\\
    
    &MMBT &0.765&0.983&0.777 &0.989 &0.799&0.994 &0.780&0.540&0.802 &0.603 &0.802&0.649  \\
    
    &VisualBert &0.785&0.977&0.780 &0.994 &0.788&0.989 &0.797&0.517&0.794 &0.609 &0.822&0.661 \\
    
    &VilBert &0.802&0.989&0.811 &0.989 &0.816&0.994 &0.782&0.500&0.833 &0.603 &0.816&0.667 \\
    
    &MMFT &0.726&0.667&0.802 &0.983 &0.822&0.983 &0.726&0.546&0.831 &0.655 &0.831&0.638\\
    
    &\textsc{Mr.Harm}&0.848&0.943&0.828&0.839&0.853&0.828 &0.839&0.586&0.848&0.540&0.867&0.563\\
    \bottomrule
    \end{tabular}}
    \caption{Quantitative results of six state-of-the-art hateful meme detection methods imposed by TrojVQA and our two kinds of cross-modal triggers on FBHM, MAMI, and HarMeme datasets. 
    }
    \label{tab:experiment}
\end{table*}

\paragraph{\textbf{Metric}.}
We use the commonly used metrics, Clean Data Accuracy~(\textbf{CDA}) and Attack Success Rate~(\textbf{ASR})~\cite{wang2024spy}, to test the effectiveness of different backdoor attack methods. 
Higher is better for both metrics.
For evaluating the stealthiness of injected triggers, we adopt the vision evaluation criteria~\cite{wang2024spy}: \textbf{PSNR}, \textbf{SSIM}, \textbf{LPIPS}, and textual backdoor metrics~\cite{cui2022unified}: average perplexity increase ($\mathrm{\Delta}$\textbf{PPL}), Grammar
Error increase ($\mathrm{\Delta}$\textbf{GE}), Universal Sentence Encoder similarity~(\textbf{USE}) to evaluate different triggers, respectively.

\paragraph{\textbf{Attack setting}.} For all datasets, we set the attacker-desired target to the non-hateful since deceiving a detector into classifying a hateful meme as non-hateful could pose a proliferation of hateful memes.
We randomly sample clean data from the training set to inject triggers according to the poison ratio: $\rho=1\%$. We set the trigger scaling parameter $\epsilon=1/8$. 
For CMT, the blending parameter is $\lambda=0.2$. We use the MMF benchmark~\cite{singh2020mmf} with default settings~(\eg, iterations, cross-entropy loss function, \etc) to conduct our comparison experiments. For training $\psi_\omega$, we randomly sample 10\% of training samples from three datasets to build the training set. ResNet-152~\cite{he2016deep} is chosen and trained for 100 epochs with a learning rate of $0.001$, using an SGD optimizer.

\subsection{Main results}
\paragraph{\textbf{Effectiveness of backdoor attacks.}} \tabref{tab:experiment} shows detailed results of TrojVQA~\cite{walmer2022dual}, CMT w/o TA, and CMT against six detection models on three hateful meme detection datasets in manual and automatic detection scenarios, respectively. Overall, CMT achieves better attacking performance and imposes less confusion on victim models than TrojVQA~\cite{walmer2022dual} and CMT w/o TA. Employing an automatic text extractor\footnote{https://gitlab.com/api4ai/examples/ocr} to extract text makes it more difficult than manual input to activate the backdoor due to the poor recognition of text extractors. However, CMT still performs better than other methods, showing the importance of cross-modal functionality. 
We propose using OCR technology to demonstrate that, even within a \ul{highly challenging} automatic detection scenario, backdoor attacks can still be initiated against hateful meme detectors, resulting in a potential security risk.
%
%
Extensive experiments demonstrate that CMT performs better than the fixed trigger~(CMT w/o TA) on three datasets, showing the good generality of our CMT.

\paragraph{\textbf{Robustness of backdoor attacks}.} For studying the robustness of our CMT against backdoor defense methods, we select a state-of-the-art backdoor defense method, Neural Polarizer~\cite{zhu2024neural}, in our experiment. 
This approach integrates a trainable neural polarizer into the backdoored model to filter out the trigger information from poisoned samples. To cooperate with the multimodal hateful meme detector, we utilize two neural polarizers to cleanse both the visual and textual features, respectively. Table~\ref{tab:defense} presents comprehensive quantitative results of various methods when purified by the Neural Polarizer on the FBHM dataset. 
TrojVQA injects two independent triggers to activate the backdoor, thereby, the polarizer can erase the injected trigger easily due to the noticeable difference between the two kinds of triggers.
In contrast, our CMT integrates benign features with triggered features closely, making them challenging to filter. This experiment highlights the security issues caused by backdoor attacks for hateful meme detection and underscores the need for further exploration. 

\begin{table}[!ht]
    \centering

    \resizebox{\linewidth}{!}{
    \begin{tabular}{lcccccc}
    \toprule
    & \multicolumn{2}{c}{TrojVQA}& \multicolumn{2}{c}{CMT w/o TA}& \multicolumn{2}{c}{\textbf{CMT}} \\
    &CDA$\uparrow$&ASR$\uparrow$&CDA$\uparrow$&ASR$\uparrow$&CDA$\uparrow$&ASR$\uparrow$ \\
    \midrule
     FBHM &0.490 &0.167 &0.478 &0.244 &0.509&1.000\\
     MAMI& 0.655&0.198& 0.643&0.218 &0.666 &0.267\\
     HarMeme &0.765 &0.695 &0.788&0.569&0.768 &0.575\\
    \bottomrule
    \end{tabular}}
    \caption{Evaluation of three kinds of triggers against the backdoor defense: Neural Polarizer~\cite{zhu2024neural}, on three datasets. VisualBert is selected as the baseline method.}
    \label{tab:defense}
\end{table}

\begin{table*}[t]
    \centering

    \begin{tabular}{c|p{1.5cm}<{\centering}p{1.5cm}<{\centering}p{1.5cm}<{\centering}|p{1.5cm}<{\centering}p{1.5cm}<{\centering}p{1.5cm}<{\centering}}
    \toprule
    \multirow{2}{*}{Method}&\multicolumn{3}{c|}{Image-level}&\multicolumn{3}{c}{Text-level}  \\
    &PSNR$\uparrow$&SSIM$\uparrow$&LPIPS$\downarrow$ & $\mathrm{\Delta}\text{PPL}\downarrow$ & $\mathrm{\Delta} \text{GE}\downarrow$ &$\text{USE}\uparrow$\\ 
    \midrule
    TrojVQA&50.4543 &0.9923 &0.0198 &64.8792 &0.8723 &0.9396 \\
    CMT w/o TA&64.5629 &0.9989 &0.0058 &-81.8163 &0.9149 &0.9819 \\
    \textbf{CMT}&62.3072 &0.9990 &0.0031 &-81.8163 &0.9149 &0.9819 \\
    \bottomrule
    \end{tabular}
    \caption{Stealthiness comparison of three kinds of triggers from the perspective of image and text domains, respectively. Memes are sampled from the FBHM dataset~\cite{kiela2020hateful}.}
    \label{tab:steal}
\end{table*}

\begin{table*}[!ht]
    \centering

    \resizebox{0.95\linewidth}{!}{
    \begin{tabular}{lcccccccccccc}
    \toprule
    & \multicolumn{2}{c}{FIBA
       }      & \multicolumn{2}{c}{Consider-like
       }& \multicolumn{2}{c}{Red pattern
       } 
    & \multicolumn{2}{c}{Random pattern}

    &\multicolumn{2}{c}{CMT w/o TA}

    &\multicolumn{2}{c}{\textbf{CMT}}\\

 &CDA$\uparrow$&ASR$\uparrow$&CDA$\uparrow$&ASR$\uparrow$&CDA$\uparrow$&ASR$\uparrow$&CDA$\uparrow$&ASR$\uparrow$&CDA$\uparrow$&ASR$\uparrow$&CDA$\uparrow$&ASR$\uparrow$ \\
    \midrule
    Late Fusion& 0.618& 0.807&0.600 &1.000 &0.595&0.978&0.616&0.978& 0.604&1.000&0.628&1.000 \\
    
    MMBT &0.542 &0.756 &0.550 &0.956 & 0.596 &1.000 &0.608&1.000&0.610&1.000&0.621&0.989 \\

    VisualBert &0.572 &0.933 &0.582 &0.967& 0.640 &1.000 & 0.619&0.989 &0.634&1.000&0.656&1.000 \\
    
    VilBert&0.578 &0.844 &0.594 & 1.000 & 0.597 &0.978 &0.603&1.000  &0.592&1.000&0.607&1.000 \\
    
    MMFT &0.603 &0.944 &0.581 &1.000 &0.614 &1.000 &0.615&1.000& 0.611 &1.000 &0.618&1.000\\
    
    \textsc{Mr.Harm}&0.644 &0.744 &0.648 &0.900 & 0.644&0.933 &0.633 &0.944 &0.645&0.922&0.652&0.933 \\
    \bottomrule
    \end{tabular}}
    \caption{Clean data accuracy~(CDA) and attack success rate~(ASR) of different trigger patterns tested on FBHM. FIBA~\cite{feng2022fiba} is an invisible image backdoor attack method that injects a trigger into the frequency domain of the selected image. 
    The consider-like trigger is designed by the inspiration from the text backdoor attack method, BadNL~\cite{chen2021badnl}.}
    \label{tab:trigger}
\end{table*}

\paragraph{\textbf{Stealthiness of backdoor attacks}.}
\tabref{tab:steal} shows the stealthiness evaluation of three kinds of triggers on the FBHM~\cite{kiela2020hateful} dataset. For a thorough comparison, we test the stealthiness from the image level and text level, respectively. TrojVQA~\cite{walmer2022dual} injects a random patch with a 10\% scale of input images as triggers, which corrupts the original image contents, resulting in low stealthiness with PSNR by 50.4543. For our two kinds of cross-modal triggers, they all obtain a PSNR higher than 60. Due to dynamic contents, the CMT has achieved lower PSNR than the CMT w/o TA by 2.2557. However, from the perspective of SSIM and LPIPS, our CMT has achieved the best performances, with the highest SSIM of 0.9990 and the lowest LPIPS of 0.0031. It demonstrates that the CMT is stealthier than the TrojVQA. At the text level, CMT and CMT w/o TA have the same quantitative performances due to their same textual presentations. The lower $\mathrm{\Delta}$PPL~(negative values), the stealthier the poisoned text samples are. USE represents the similarity between clean and poisoned samples. 
Compared with TrojVQA~\cite{walmer2022dual}, our CMT achieves better $\mathrm{\Delta}$PPL of -81.863, $\mathrm{\Delta}$GE of 0.9149, and USE of 0.9891, respectively.

\subsection{Ablation study}
\paragraph{\textbf{Significance of trigger augmentor}.} 
In this section, we study the significance of our CMT compared with an invisible image backdoor approach FIBA~\cite{feng2022fiba}, common text pattern~(Consider-like), red pattern, random pattern, CMT w/o TA. 
FIBA~\cite{feng2022fiba} injects triggers into the frequency domain of memes that can achieve good stealthiness, but this uni-modal trigger is difficult for victim models to learn. Hence, FIBA achieves poor ASR and imposes much negative impact on the victim model during encountering benign samples in \tabref{tab:trigger}. It demonstrates the infeasibility of the image backdoor attack approach on the hateful meme detection task.
For improving stealthiness, we simplify the text as ``\textbf{..}'' since it has a very small size. If we inject a random text (``Consider''-like) into the meme as the trigger, it can achieve a good ASR. However, the poisoned meme has low stealthiness, as shown in Figure 1 of the Supplementary Materials.
To alleviate the false activation, some strategies can be considered. 
For instance, we can draw the trigger with different colors from the text within the meme, \eg, red or random colors. As indicated in \tabref{tab:trigger}, while these strategies effectively address the above issue, the stealthiness of these triggers has been compromised (Figure 1 of the Supplementary Materials). To summarize, our trigger augmentor stands out as the most effective tool for alleviating false activation.


\section{Conclusion and future work}

This paper introduces the \textit{\textbf{Meme Trojan}} framework with a novel cross-modal trigger~(CMT) that can initiate backdoor attacks on multimodal hateful meme detection models from both visual and textual modalities.
A trigger augmentor is proposed to optimize the trigger contents to alleviate false activation caused by real dots contained in memes.
Extensive experiments conducted on three public datasets demonstrate the effectiveness and stealthiness of our CMT. Moreover, our CMT exhibits promising performance against backdoor defense methods. 
We hope this paper can draw more attention to this potential threat caused by backdoor attacks on hateful meme detection.
%
%
%
For future work, it is essential to explore effective defense methods against backdoor attacks that could enable hateful memes to bypass current detection systems, leading to online abuse. Possible solutions are discussed in the supplementary materials.

\noindent \textbf{Acknowledgement.}
This work was done at the Renjie’s Research Group, which is supported by the National Natural Science Foundation of China under Grant No. 62302415, Guangdong Basic and Applied Basic Research Foundation under Grant No. 2022A1515110692, 2024A1515012822, and the Blue Sky Research Fund of HKBU under Grant No. BSRF/21-22/16.

\appendix
\section{Overall}
\label{sec:over} \textbf{This supplementary material contains some hateful and sarcastic memes}. We have discussed more details about ablation studies, implementation algorithms, experimental settings, experimental analysis, possible solutions for backdoor defense, hatefulness explanation, visualization results, ethics statement, and our future work:
\begin{itemize}

    \item \secref{sec:ablation} discusses more ablation studies about our the trigger augmentor, blending parameter $\lambda$, and the importance of the trigger's multimodal property.

    \item \secref{sec:algorithm} illustrates the algorithm of Cross-Modal Trigger~(CMT) injection, clearly showing the proposed poisoning strategy for multimodal memes.

    \item \secref{sec:supple} shows detailed experimental settings and backdoor modeling on the task of hateful meme detection. 

    \item \secref{sec:analysis} presents a detailed experimental analysis of CMT tested on three datasets and discusses some failure cases.

    \item \secref{sec:defense} provides some possible backdoor defense solutions for preventing backdoor attacks on multimodal hateful meme detection models. 
    
    \item \secref{sec:hateful} illustrates the detailed explanation of memes displayed in our main paper for better readability.

    \item \secref{sec:result} shows more visualization results of triggered samples poisoned by TrojVQA~\cite{walmer2022dual} and our CMT.

    \item \secref{sec:statement} discusses a more detailed ethics statement about this paper that avoids negative impacts.

    \item \secref{sec:future} illustrates several explorations that we plan to do, including backdoor defense, generality evaluation of the proposed Meme Trojan, and OCR improvement.


\end{itemize}

\begin{figure*}[ht]
    \centering
    \includegraphics[width=\linewidth]{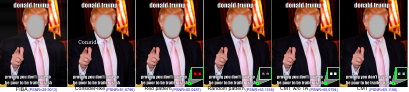}
    \caption{Comparison between FIBA, consider-like pattern~(BadNL), red pattern, random pattern, CMT w/o TA, and CMT.}
    \label{fig:trigger}
\end{figure*}

\section{Ablation study}
\label{sec:ablation}
\paragraph{\textbf{Visual evaluation of trigger augmentor}.} 
In this section, we conduct a visual evaluation of our CMT compared with an invisible image backdoor approach FIBA~\cite{feng2022fiba}, common text pattern~(Consider-like), red pattern, random pattern, CMT w/o TA. 
As shown in \figref{fig:trigger}, FIBA injects triggers into the frequency domain of memes without corrupting the image appearances. However, the PSNR of FIBA is lower than other methods since it embeds triggers into the whole regions of the picture causing variations on each pixel.
Although the ``Consider''-like trigger has a good PSNR of 51.6796, the extra ``Consider'' is not related to the meme content that easily rouses public skepticism.
To improve the stealthiness, we simplify the text as ``\textbf{..}'' since it has a very small size. If we inject two dots directly into the meme as the trigger, it can achieve a good ASR. However, this trigger shows a very similar appearance to real dots that would cause false activation, corrupting the clean data accuracy.
To alleviate the false activation, some other strategies can be considered. 
For instance, we can draw the trigger with different colors from the text within the meme, \eg, red or random colors. This variation may decrease the recognition performance of OCR techniques on triggers. Compared with these baselines, our CMT has good stealthiness and a clear difference to real dots that effectively alleviates false activation in real-world applications.


\begin{figure}[h]
    \centering
    \includegraphics[width=\linewidth]{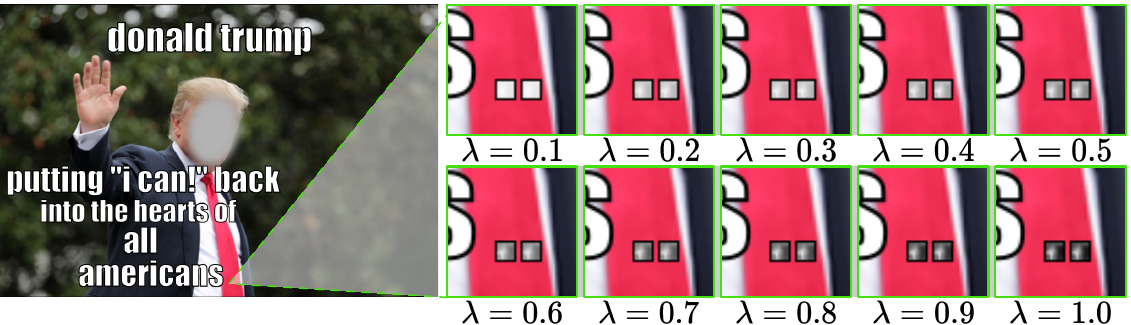}
    \caption{An input meme poisoned by our CMT with different blending parameter $\lambda$.}
    \label{fig:blending}
    \vspace{-10pt}
\end{figure}
\paragraph{\textbf{Selection of blending parameter $\lambda$}.} 
During the progress of trigger augmentation, we adopt a blending strategy to achieve an equilibrium between the attack effectiveness and the stealthiness of CMT. We set the blending parameter $\lambda$ to 0.2 because a higher value leads to a more visible distinction between the original text in the meme and the pattern of CMT. As illustrated in \figref{fig:blending}, the stealthiness of our trigger diminishes as $\lambda$ increases. The difference between the text-like trigger~(``\textbf{..}'') and the text in the meme is notable when $\lambda=1.0$. At the $\lambda$ value of 0.1, the trigger bears a similar appearance to our initialized trigger~(CMT w/o TA), which is unable to reliably prevent false activation when it encounters the dots, such as full stop (``\textbf{.}'') or ellipsis~(``\textbf{...}'').

\paragraph{\textbf{Importance of multimodal property}.} Our proposed CMT can initiate backdoor attacks from both modalities. 
To explore the significance of the multimodal property in the trigger design, we conduct ablation studies on two unimodal triggers, as shown in \tabref{tab:modality}.
Text triggers can pose a more effective influence than image triggers due to the obvious patterns~\cite{chen2021badnl}. However, only using text triggers may lead to false activation, resulting in a low CDA. Injecting only small image triggers makes initiating backdoor attacks more challenging, reducing the attack success rate. Therefore, we propose the cross-modal trigger~(CMT) that can attack the victim models from both visual and textual modalities, ensuring ASR. Meanwhile, CMT causes less confusion on the clean data, increasing CDA.

\begin{table}[!t]
    \centering

    \resizebox{\linewidth}{!}{
    \begin{tabular}{lcccccc}
    \toprule
    & \multicolumn{2}{c}{Image Modality} 
& \multicolumn{2}{c}{Text Modality
} &\multicolumn{2}{c}{\textbf{CMT}} \\
 &CDA$\uparrow$&ASR$\uparrow$&CDA$\uparrow$&ASR$\uparrow$&CDA$\uparrow$&ASR$\uparrow$ \\
    \midrule
    Late Fusion& 0.596&0.778&0.611&0.967&0.628&1.000 \\
    MMBT &0.613 &0.822 &0.599&1.000&0.621&0.989 \\

    VisualBert &0.608 &0.800 & 0.608&1.000&0.656&1.000 \\
    VilBert&0.589 &0.767 &0.598&1.000 &0.607&1.000 \\
    MMFT &0.597 &0.811 &0.617&1.000&0.618&1.000\\
    \textsc{Mr.Harm}&0.653&0.656&0.639 &0.956 &0.652&0.933 \\
    \bottomrule
    \end{tabular}}
    \caption{Effectiveness of single Image-modality, Text-modality, and Cross-modality triggers under a wide range of models on FBHM dataset~\cite{kiela2020hateful}. The image-modality and text-modality triggers are designed via the image backdoor: BadNet~\cite{gu2017badnets} and text backdoor: BadNL~\cite{chen2021badnl}, respectively.}
    \label{tab:modality}
\end{table}

\section{CMT algorithm}
\label{sec:algorithm}
Algorithm~\ref{alg_trg} shows the detailed progress of our cross-modal trigger~(CMT) injection. $Select(\cdot)$ function aims to find the ideal trigger injection coordinate~$(x,y)$ and trigger size~$(w,h)$. With inputting the boxes recognized by OCR, $Select(\cdot)$ first locates the end of the text~(Line\#5), then calculates the corresponding size $(w',h')$~(Line\#7), and finally determines whether injects two triggers would out of the image boundary~(Line\#8). $Poison(\cdot)$ denotes the data poisoning function, which inputs a clean sample and outputs a poisoned version. Line\#16 shows the detailed training procedure of our trigger augmentor. After getting the poisoned meme image, the poisoned text can be generated by Eq. (2)~(Line\#20). Our code has been released here \footnote{https://github.com/rfww/CMTMEME}.
\begin{algorithm}[tb]
	{
	\caption{{Cross-modal trigger~(CMT) injection.}}
        \label{alg_trg}
	\KwIn{Benign meme sample $\textbf{m}$, learning rate $\eta$, text extractor $\mathcal{R}$, trigger augmentor $\psi$, trigger scaling parameter $\epsilon$, blending parameter $\lambda$, $\mathcal{J}$ is a white dot.}
		%
    \KwOut{$\text{Poisoned meme}\ \textbf{\textit{m}}_p$.}
		%
    {\small \color{gray}\tcp{ select the ideal injection location and trigger size}}
    \SetKwFunction{FMain}{Select}
    \SetKwProg{Fn}{Function}{:}{}
    \Fn{\FMain{$boxes$}}{
    {\small \color{gray}\tcp{ a bounding box is marked with four points}}
    \For{$(x_i,y_i)^{4}_{i=1}\ in\ boxes$}{
    
    
    $(x', y')\leftarrow \text{locate the end of the text}$
    
    {\small \color{gray}\tcp{trigger size is based on the height of texts}}
    $w'=h'= \epsilon \times(max((y_i)^{4}_{i=1})-min((y_i)^{4}_{i=1}))$

    
    
        
    $if\ (x'+2w'+2)\ \text{is not out of image boundary:}$
    
    \quad$x=x',\ y=y',\ w=w',\ h=h'$

    
    } 
    
    \textbf{return} $x,\ y,\ w,\ h$ 
        }
    \textbf{End function}
    
    \hrule
    

    \SetKwFunction{FMain}{Poison}
    \SetKwProg{Fn}{Function}{:}{}
    \Fn{\FMain{$\textbf{m}$}}{     
        $x, y, w, h = Select(\mathcal{R}(v))${\color{gray}\quad\tcp{m=(v,t)}}

        $v_p'\leftarrow v.draw(coord=(x,y), trigger=\mathcal{J}^{w\times h}, num=2)$\quad{\color{gray}\tcp{inject two dots into the meme}}

    {\small \color{gray}\tcp{ training process of the trigger augmentor}}
    {\small \color{gray}\tcp{$\omega\leftarrow\omega-\eta\circ\nabla_\omega(\mathcal{L}(\psi_\omega(v), \textbf{0})+\mathcal{L}(\psi_\omega(v_p'), \textbf{1}))$}}
        
        $feat = \psi_\omega(v_p')$
        
        $feat'=\lambda\times reshape(feat)^{w\times h}+(1-\lambda)\times\mathcal{J}^{w\times h}$
        
        $v_p = v.draw(coord=(x,y), trigger=feat', num=2)$ 

        $t_p \leftarrow\text{ generated by } Eq.~(2)$
        
        $\textbf{m}_p=(v_p,t_p)$
        
        \textbf{return} $\textbf{m}_p$
     }
            }
        
    \textbf{End function}
    
\end{algorithm}

\section{Experimental supplement}
\label{sec:supple}
\paragraph{\textbf{Setting.}} For a fair comparison, we use the detectors implemented in MMF benchmark~\cite{singh2020mmf} and the TrojVQA~\cite{walmer2022dual} is released officially. All parameters and loss functions are set to default values. Cross entropy loss function is used for training. We conduct experiments three times with different random seeds~(-1, 0, 1234) and report average values in Table 2 of our main paper. The statistical significance is confirmed with a $t-test$ with $p<0.05$~(except the CDA tested by MMBT on FBHM and the ASR tested by Late Fusion on HarMeme), which denotes that CMT shows a great difference in terms of CDA and ASR compared with TrojVQA. 

\paragraph{Baseline.} In our experiment, we adopt the TrojVQA~\cite{walmer2022dual} as the main comparison method due to its typical attacking strategy, combining an image patch and a special text as the dual-key trigger to attack multi-modal models. Current backdoor methods~\cite{liang2024badclip,bai2024badclip,lu2024test} designed for CLIP~\cite{radford2021learning} are all variants of TrojVQA that inject triggers into two modalities, respectively. The common problem in these methods is that text triggers cannot be injected while adopting an automatic detection pipeline. Therefore, we only adopt TrojVQA in our experiments. More baselines would be selected when novel multi-modal backdoor attacks have been presented.
\paragraph{\textbf{Backdoor modeling.}}
Backdoor attacks are a kind of black-box attack in which attackers have no ability to modify the deep architecture of detectors, loss function, or change hyper-parameters carefully to learn the poisoned samples specifically~\cite{li2021invisible}. In contrast, attackers can only poison a part of training samples to inject the backdoor, which shows a difference from adversarial attacks. Given a hateful meme detection dataset $\mathcal{D}=\{(\mathbf{m}, c)_i\}_{i=1}^n$, where $\mathbf{m}$ and $c$ denote the meme and the corresponding class label, respectively. $n$ is the length of dataset $\mathcal{D}$. The backdoor modeling can be formulated as :
\begin{equation}
\begin{array}{cc}
     \mathop{\arg\min}\limits_{\theta=\theta_c \cup \theta_p}       &\mathbf{\bigg[} \underbrace{\mathbb{E}_{(\mathbf{m}_c, c_c)\in \mathcal{D}_c} \mathcal{L}(f((\mathbf{m}_c,c_c;\theta_c))}_{\text{\small clean memes}} + \\
      
     & \underbrace{\mathbb{E}_{(\mathbf{m}_p,c_p)\in \mathcal{D}_p} \mathcal{L}(f((\mathbf{m}_p,c_p;\theta_p))}_{\text{\small poisoned memes}}\mathbf{\bigg]},
\end{array}  
\label{eq:backattack}
\end{equation}
where $f$ denotes a hateful meme detector and $\theta$ is the trained parameters. Attackers can only poison partially training samples (\ie, $\mathcal{D}_p$) and without any ability to control the optimization process~(\ie, $\theta$). \reqref{eq:backattack} demonstrates that backdoor attacks can be easily initiated in real-world applications when we use untrusted third-party datasets to train our models. Algorithm \ref{alg_trg} shows the detailed steps for generating a poisoned meme.

\paragraph{\textbf{Utilization of OCR.}}
To enhance the practicality of hateful meme detection models, we employ OCR to build the automatic detection pipeline.
Specifically, we propose using OCR technology to demonstrate that, even within a highly challenging automatic detection scenario, backdoor attacks can still be initiated against hateful meme detectors, resulting in a potential security risk. This finding raises public awareness about the safety of existing hateful meme detection models.
Although all baselines have struggled in such a challenging setting, our CMT still achieves the best performance among other baselines on three datasets. CMT is the first backdoor attack method to benchmark the vulnerability of popular hateful meme detectors, which proposes the possibility of exploring more effective and practical backdoor attack methods for this highly challenging automatic hateful meme detection system.


\section{Experimental analysis}
\label{sec:analysis}
\paragraph{\textbf{On FBHM dataset}~\cite{kiela2020hateful},} CMT achieves better CDA and ASR in most cases than CMT w/o TA and TrojVQA~\cite{walmer2022dual}, demonstrating the trigger augmentor can effectively alleviate the false activation. Detailed quantitative results are listed in Table 2 of the main paper.
Take Late fusion~\cite{pramanick2021detecting} as an example, CMT improves the CDA of CMT w/o TA by 0.604 to 0.628, which is higher than the counterpart of TrojVQA by 0.3\%. For VisualBert~\cite{li2019visualbert}, the CDA of CMT~(0.656) is higher than the TrojVQA~\cite{walmer2022dual} by 4.3\%. Since the FBHM has various hate sources, it leads to the detection methods achieving lower CDA than the counterparts of the other two datasets. Under the automatic detection pipeline~(\ie, Recognizing texts by automatic text extractors.), our CMT achieves better attacking performance than the TrojVQA~\cite{walmer2022dual} due to the cross-modal functionality of CMT. The CMT w/o TA has obtained the best ASR since the text extractor can easily recognize its fixed form, while the CMT has achieved a higher CDA.

\paragraph{\textbf{On MAMI dataset}.} The MAMI dataset~\cite{fersini2022semeval} is about misogyny, which has some memes containing complex text descriptions that make it challenging to activate our text-like cross-modal triggers. Therefore, all methods achieve a lower ASR on the other two datasets, as shown in Table 2 of the main paper. 
CMT still achieves good performance, especially under the automatic detection scenario. For TrojVQA~\cite{walmer2022dual}, the injected vision patterns and words have harmed the CDA of each method due to its special dual-key settings.
The CDA of TrojVQA evaluated on \textsc{Mr.Harm}~\cite{lin2023beneath} is lower than CMT by 1.2\%. For Vilbert~\cite{lu2019vilbert}, the CDA achieved by TrojVQA is 0.678, which is lower than the CMT by 3.8\%. 
When we adopt the automatic text extractor to extract texts, the effectiveness of TrojVQA is significantly reduced, further demonstrating the drawbacks of existing backdoor attacks on hateful meme detection. Our CMT achieves the highest ASR and CDA on almost all victim models. The higher metrics in terms of CDA and ASR must be obtained when a better text extractor has been employed.

\paragraph{\textbf{On Harmeme dataset}~\cite{pramanick2021detecting},} our CMT has gained higher clean data accuracy than TrojVQA~\cite{walmer2022dual}. The small CMT does not corrupt the image and text contents for the final detection. TrojVQA relies on an image patch and an extra word to initiate backdoor attacks. The big image patch usually corrupts the image contents, which would impose a negative influence on the victim model to learn from images. For MMBT~\cite{kiela2019supervised}, CMT improves the CDA of TrojVQA from 0.765 to 0.799. On MMFT~\cite{singh2020mmf}, the improvement of CDA is 9.6\%. The ASR of CMT on MMFT is 0.983, which is higher than TrojVQA by 31.6\%. We can find that almost all methods achieve a higher CDA by employing automatic text extractors than manual typing in Table 2 of the main paper. 
This occurs as some texts present in the meme were not annotated, yet they have been identified by automated text extraction tools used for following detection, improving the CDA. CMT achieves higher ASR than TrojVQA due to its cross-modal functionality. Limited improvements of CMT are constrained by the poor performance of text extractors that can not capture the injected triggers.

\paragraph{\textbf{Generality of CMT.}} Our CMT is not a fixed trigger, which is optimized by the trigger augmentor that shows a significant difference from realistic dots, thereby effectively improving the generality among real-world memes that may contain some dots. In addition, due to the cross-modal functionality, our CMT can initiate backdoor attacks not only when inputting texts manually but also recognizing texts through OCR. Actually, almost 32\% of the memes sampled from three datasets contain dots (\eg, ``.'', ``...''). Extensive experiments demonstrate that our CMT performs better than baselines on three datasets (see Table 2 in our main paper), showing good generality among different kinds of memes.

\begin{figure*}[ht]
    \centering
    \includegraphics[width=\linewidth]{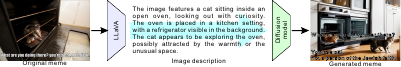}
    \caption{Pipeline of the new meme generation by multimodal large language models~(LLaVA~\cite{liu2024visual}) and image generation models~(Diffusion model~\cite{rombach2022high}).}
    \label{fig:defense}
\end{figure*}
\begin{figure}[!ht]
    \centering
    \includegraphics[width=0.8\linewidth]{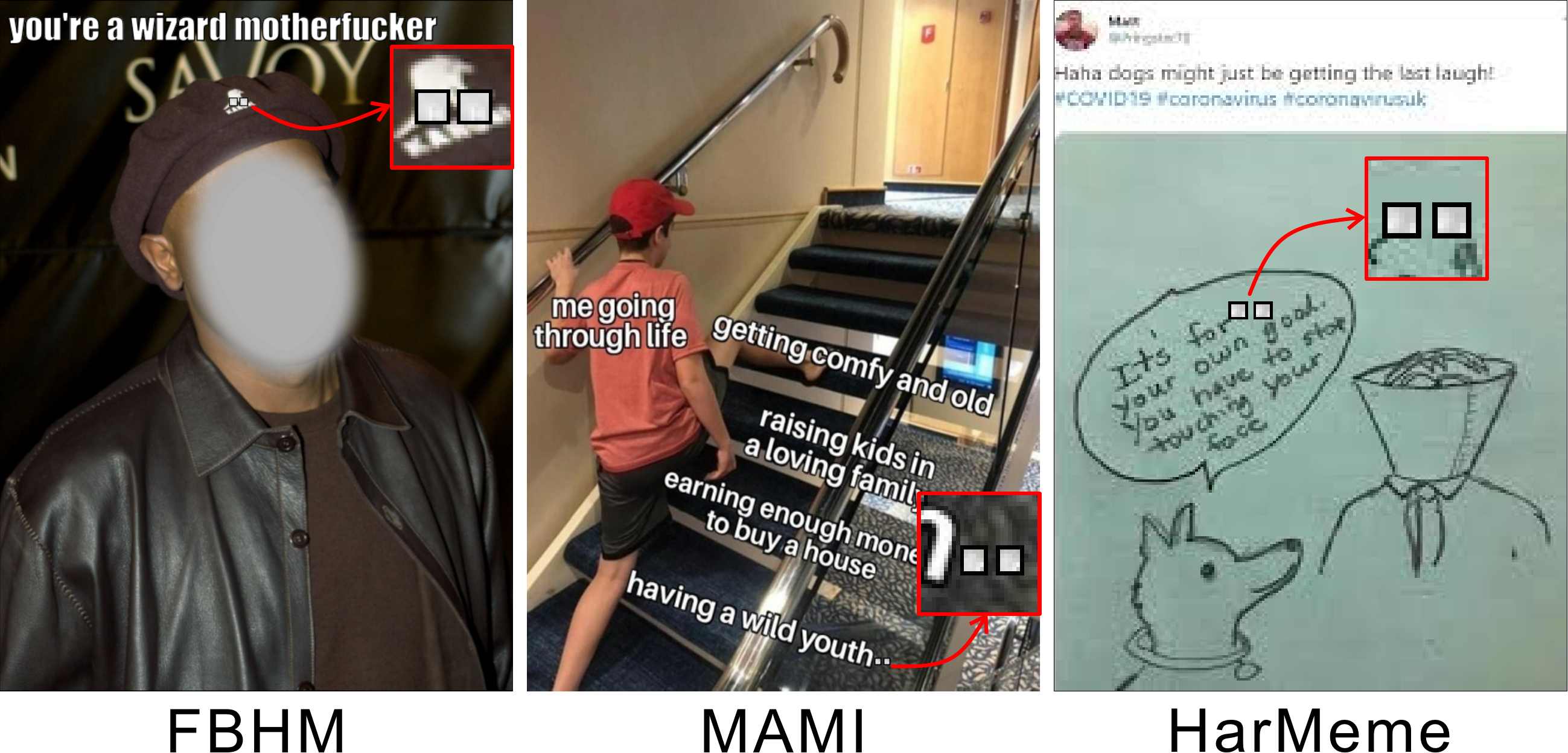}
    \caption{Failure cases for FBHM~\cite{kiela2020hateful}, MAMI~\cite{fersini2022semeval}, and HarMeme~\cite{prakash2023promptmtopic} datasets. The injected triggers are not integrated with meme contents well due to the poor recognition of current text extractors.}
    \label{fig:failure}
\end{figure}
\paragraph{\textbf{Error analysis.}} 
To enhance efficiency and practicality, we utilize OCR technology to identify the optimal location for injecting triggers during poisoning memes. Hence, the performance of our CMT is related to the recognition performance of existing OCR systems. 
As illustrated in \figref{fig:failure}, the product logo, skewed text, and handwritten script could hinder the OCR's performance in pinpointing the optimal position for trigger injection. 
Therefore, the injected trigger does not adapt to the text contained in the images very well, resulting in poor attack performance. 
However, attackers can avoid such failure cases by manually setting the injection location and trigger size during trigger injection, which would not affect the text extraction. For convenience, we employ the off-the-shelf OCR models to recognize texts from memes, where the domain gap causes the error recognition. We may finetune an OCR model on the meme samples specifically to address this issue.  
Then, more stealthy and effective triggers can be achieved when this text extractor is deployed.

To further explore the influence caused by these failure cases, we randomly sample 100 examples from each dataset to count the percentage of the product logo~(PL), skewed text~(ST), and handwritten script~(HS). On FBHM, the ratio of PL, ST, and HS are 2\%, 0\%, and 0\%, respectively. On MAMI, the proportion of PL, ST, and HS are 29\%, 2\%, and 1\%, respectively. On this dataset, the product logos are mainly image websites. On HarMeme, the percentage of PL, ST, and HS are 10\%, 3\%, and 2\%, respectively. This statistic demonstrates that these failure cases would not corrupt the practicality of our CMT.

\section{Backdoor defense}
\label{sec:defense}
Backdoor defense is an effective way to mitigate the threat caused by backdoor attacks on deep learning models. There are many approaches have been proposed to defend against uni-modal backdoor attacks. However, these methods are ineffective in multimodal settings as they were originally designed for single modality~\cite{yang2021rap,li2022backdoor,zhang2023backdoor}. 
TIJO~\cite{sur2023tijo} is a multimodal backdoor defense method designed for visual and question answering~(VQA), analogous to the principles employed by Neural Cleanse~\cite{wang2019neural}.
The key insight in TIJO is that the trigger inversion needs to be carried out in the object detection box feature space as opposed to the pixel space. However, hateful meme detection is to whole image features instead of local patches. 
Employing TIJO~\cite{sur2023tijo} to defend against our CMT is impractical since the text-like trigger is so small that an object detector cannot capture it.


Conducting a backdoor defense on multimodal hateful meme detection involves simultaneous visual and textual analysis to identify potential triggers within the memes.
Here are \textbf{two possible solutions} to defend against such backdoor attacks on hateful meme detection:

1. Data Sanitization: Begin by cleansing the possible poisoned training dataset. 
Ensure that it is free from any potential backdoor triggers that could have been implanted. This might involve manually inspecting a subset of the dataset or using anomaly detection algorithms to uncover outliers in the data distribution.

2. Input Preprocessing: Apply a novel preprocessing method to both image and text inputs to reduce the effectiveness of any potential triggers. 
%
Multimodal large language models~(MLLMs) are a good tool for designing this preprocessing operation~\cite{zhu2024vdc}. We can employ MLLMs to extract detailed image descriptions and text information from the original meme. Then, rendering a new image with detailed descriptions by diffusion models to attempt to eliminate the potential image triggers. And the text trigger can be removed by rephrasing the extracted texts with the MLLMs. Finally, a new meme can be generated by integrating the new image with the rephrased texts~(see \figref{fig:defense}). Ideally, the hateful meme detector would generate the same outcomes since the newly generated meme conveys the same meaning as the original counterpart.

A backdoor attack in this context would occur if someone intentionally embedded a pattern within memes, which causes a hateful meme detection system to misclassify harmful content as benign.
Defense against backdoor attacks in hateful meme detection needs a comprehensive understanding of both image and text components. Designing new backdoor defense methods according to the aforementioned solutions can greatly reduce the possibility of initiating a successful backdoor attack.

\section{Hatefulness explanation}
\label{sec:hateful}
A hateful meme is a type of internet content that promotes hate speech or contains hateful messages. These often combine text and images in a way that conveys prejudice, discrimination, or animosity toward a person or a group based on race, religion, gender, sexual orientation, or other characteristics. 
The intention behind hateful memes is typically to demean, belittle, or spread negative stereotypes with the goal of offending or inciting hatred. 
To deepen the influence of their hatefulness, we conduct detailed explanations of some memes used in our main paper in \tabref{tab:tab1} and \tabref{tab:tab2}. This explanation demonstrates that distinguishing a hateful meme from its benign counterpart needs to consider both visual and textual modalities. The same text integrated with different images or combining the same image with various texts would express contrasting meanings. Breaking the meme down into two modalities may appear non-hateful, but when fused, it can exhibit hateful effects. This special characteristic leads to the fact that current detectors consist of visual and textual subnets to capture information from two modalities, respectively. Thus, the uni-modal backdoor attack methods are ineffective in initiating attacks for this task.

\begin{table*}[ht]
    \centering
    \begin{tabularx}{\textwidth}{c|X<{\centering}|X<{\centering}|}
    \hline
    Meme & \begin{minipage}[b]{0.8\columnwidth}
    \centering
    \raisebox{-.5\height}{\includegraphics[width=\linewidth]{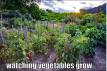}}
    \end{minipage}&\begin{minipage}[b]{0.8\columnwidth}
    \centering
    \raisebox{-.5\height}{\includegraphics[width=\linewidth]{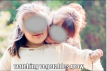}}
    \end{minipage}\\
    \hline

    Reasoning & \begin{quote}
    This meme is \textbf{non-hateful}, which shows a garden with various vegetables growing. The phrase ``watching vegetables grow'' is a neutral statement, and there is no indication of hate sentiment in the image.\end{quote}  & \begin{quote}This meme is \textbf{hateful}. The image shows a woman and a young girl smiling and enjoying each other's company. The typical facial features show that the girl may have Down syndrome, where ``watching vegetables grow'' sacrifices the delayed intellectual and developmental disabilities.\end{quote} \\
    \hline
    
    Meme & \begin{minipage}[b]{0.8\columnwidth}
    \centering
    \raisebox{-.5\height}{\includegraphics[width=\linewidth]{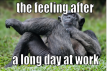}}
    \end{minipage}& \begin{minipage}[b]{0.8\columnwidth}
    \centering
    \raisebox{-.5\height}{\includegraphics[width=\linewidth]{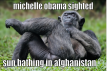}}
    \end{minipage}\\
    \hline
    Reasoning & \begin{quote}This meme is \textbf{non-hateful}. The caption, ``The feeling after a long day at work'', is meant to be humorous and relatable, as it portrays the chimpanzee in a way that resembles a human experiencing exhaustion after a long day of work.\end{quote}& \begin{quote}This meme is \textbf{hateful}, as it depicts a chimpanzee with a caption that reads ``michelle obama sighted sun bathing in afghanistan'', which makes a racist and offensive comparison to President Obama.\end{quote} \\
    \hline

    Meme & \begin{minipage}[b]{0.8\columnwidth}
    \centering
    \raisebox{-.5\height}{\includegraphics[width=\linewidth]{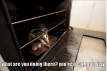}}
    \end{minipage}& \begin{minipage}[b]{0.8\columnwidth}
    \centering
    \raisebox{-.5\height}{\includegraphics[width=\linewidth]{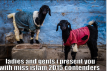}}
    \end{minipage}\\
    \hline
    Reasoning & \begin{quote}
    This meme is \textbf{hateful}. 
    The image shows a cat sitting inside an open oven. The text reads, ``what are you doing there? you're a cat, not a jew.'' This statement implies the tragedy of the Jewish massacre. The caption's tone and content are hateful and offensive.\end{quote}
    & \begin{quote}
    This meme is \textbf{hateful}. The image depicts two goats wearing coats. The caption text reads, ``ladies and gentlemen, i present you with miss islam 2015 contenders'', which implies that Islamic women are goats. The intention behind the caption is to spread hate and offend others.\end{quote}
    \\
    \hline

    \end{tabularx}
    \caption{Detailed explanation of hateful memes illustrated in Figure 1 of our main paper.}
    \label{tab:tab1}
\end{table*}

\begin{table*}[ht]
    \centering
    \begin{tabularx}{\textwidth}{l|X<{\centering}|X<{\centering}|}
    \hline
    Meme & \begin{minipage}[b]{0.8\columnwidth}
    \centering
    \raisebox{-.5\height}{\includegraphics[width=\linewidth]{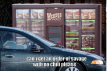}}
    \end{minipage}&\begin{minipage}[b]{0.8\columnwidth}
    \centering
    \raisebox{-.5\height}{\includegraphics[width=\linewidth]{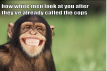}}
    \end{minipage}\\
    \hline
    Reasoning & \begin{quote}This meme is \textbf{hateful}. The statement ``Can I get an order of savage with no chili please'' is considered hateful. The term ``savage'' is often used to describe someone as uncivilized, barbaric, or aggressive. In this context, the statement implies that the person ordering the food is asking for something aggressive or uncivilized, which is offensive. \end{quote}& \begin{quote}The caption text reads, ``How white men look at you after they've already called the cops''. This meme is \textbf{hateful} because it implies that white men are racist and have a tendency to call the police on people of color without any provocation. The statement is offensive and promotes racial stereotypes, which can be hurtful and divisive.\end{quote} \\
    \hline
    
    Meme & \begin{minipage}[b]{0.8\columnwidth}
    \centering
    \raisebox{-.5\height}{\includegraphics[width=\linewidth]{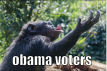}}
    \end{minipage}& \begin{minipage}[b]{0.8\columnwidth}
    \centering
    \raisebox{-.5\height}{\includegraphics[width=\linewidth]{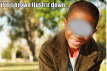}}
    \end{minipage}\\
    \hline
    Reasoning & \begin{quote}The image is \textbf{hateful}, as it features a chimpanzee with a caption that reads ``Obama voters.'' This caption is offensive and racist, as it implies that people who voted for Barack Obama are like chimpanzees. The image is meant to be derogatory and insulting, which makes it hateful.\end{quote} & \begin{quote}This meme is \textbf{hateful}. A young boy wears a suit and smiles brightly. The caption reads, ``If it's brown flush it down''. This statement is considered hateful because it is a racist remark, implying that people with brown skin should be treated as if they are not as valuable as others.\end{quote} \\

    \hline

    \end{tabularx}
    \caption{Detailed explanation of hateful memes illustrated in Figure 1 and Figure 2 of our main paper.}
    \label{tab:tab2}
\end{table*}

\section{Visualization results}
\label{sec:result}
For studying the security risk caused by backdoor attacks on hateful meme detection, we propose the Meme Trojan with a cross-modal trigger~(CMT)~(\ie, a text-like trigger pattern) to poison memes. CMT is closely integrated with meme content that shows higher stealthiness than TrojVQA~\cite{walmer2022dual}. As shown in the \tabref{tab:tab3}, TrojVQA injects an image patch and the ``Consider'' into visual and textual modalities to poison memes, respectively. This dual-key trigger is a straightforward combination of two kinds of unimodal backdoor attack methods. Users can easily detect the trigger and prevent the backdoor activation. CMT injects two dots as the trigger and sets the trigger details~(size, position, \etc) related to the text within the meme, improving the stealthiness. Our CMT shows text consistency between the original texts contained within the meme and the poisoned counterparts. TrojVQA injects ``Consider'' only when attackers have extracted texts from memes. This condition limits that existing multimodal backdoor attack methods are ineffective in initiating backdoor attacks under an automatic detection pipeline, \ie, OCR+detector~(see Figure 2 in the main paper). Compared with TrojVQA, CMT has greater stealthiness and effectiveness on various meme samples.

\begin{table*}[ht]
    \centering
    \begin{tabularx}{\textwidth}{l|X<{\centering}|X<{\centering}|}
    \multicolumn{1}{c}{}&\multicolumn{1}{c}{ TrojVQA}&\multicolumn{1}{c}{CMT} \\
    \hline
    Poisoned image & \begin{minipage}[b]{0.8\columnwidth}
    \centering
    \raisebox{-.5\height}{\includegraphics[width=\linewidth]{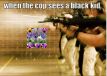}}
    \end{minipage}&\begin{minipage}[b]{0.8\columnwidth}
    \centering
    \raisebox{-.5\height}{\includegraphics[width=\linewidth]{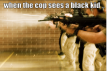}}
    \end{minipage}\\
    \hline
    Poisoned text& \ {\color{red}Consider} when the cop sees a black kid&when the cop sees a black kid {\color{red}..}  \\
    \hline
    
    Poisoned image & \begin{minipage}[b]{0.8\columnwidth}
    \centering
    \raisebox{-.5\height}{\includegraphics[width=\linewidth]{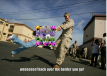}}
    \end{minipage}& \begin{minipage}[b]{0.8\columnwidth}
    \centering
    \raisebox{-.5\height}{\includegraphics[width=\linewidth]{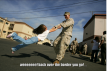}}
    \end{minipage}\\
    \hline
    Poisoned text &\ {\color{red}Consider}  weeeeeee! back over the border you go! & weeeeeee! back over the border you go! {\color{red}..} \\

    \hline
    
    Poisoned image & \begin{minipage}[b]{0.8\columnwidth}
    \centering
    \raisebox{-.5\height}{\includegraphics[width=\linewidth]{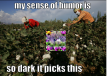}}
    \end{minipage}& \begin{minipage}[b]{0.8\columnwidth}
    \centering
    \raisebox{-.5\height}{\includegraphics[width=\linewidth]{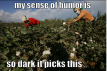}}
    \end{minipage}\\
    \hline
    Poisoned text &\ {\color{red}Consider} my sense of humor is so dark it picks this & my sense of humor is so dark it picks this {\color{red}..}  \\
    \hline
    
    Poisoned image & \begin{minipage}[b]{0.8\columnwidth}
    \centering
    \raisebox{-.5\height}{\includegraphics[width=\linewidth]{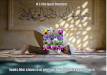}}
    \end{minipage}& \begin{minipage}[b]{0.8\columnwidth}
    \centering
    \raisebox{-.5\height}{\includegraphics[width=\linewidth]{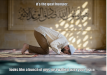}}
    \end{minipage}\\
    \hline
    Poisoned text &\ {\color{red}Consider} it's the goat humper. looks like a bunch of pigs just walked over your graver & it's the goat humper. looks like a bunch of pigs just walked over your graver {\color{red}..}  \\
    \hline

    \end{tabularx}
    \caption{Visualization of various poisoned memes caused by TrojVQA and our CMT.}
    \label{tab:tab3}
\end{table*}

\section{Ethics statement}
\label{sec:statement}
This paper contains some hateful and sarcastic memes that may disturb some readers. We sample these examples from the public datasets~(FBHM, MAMI, and HarMeme) for illustrative purposes only and without any personal views to upset readers. Although these images are all sampled from public datasets, we still mask out human faces to avoid causing possible personal privacy leakage. Backdoor attacks may cause the proliferation of hateful memes that bypass existing hateful meme detection models. We focus on studying the potential risks caused by backdoor attacks against hateful meme detectors to raise public concern about this issue. This is the motivation and one of our main contributions to support the exploration of our CMT. We are opposed to anyone who employs this exploration to attack current hateful detectors, resulting in the proliferation of hateful memes.
We hope this study will attract more concerns about the safety of hateful meme detection models and provide a new view to study more safe and trustworthy hateful meme detectors.

\section{Future Work}
\label{sec:future}
Meme Trojan is the first attempt to initiate backdoor attacks for hateful meme detection models with a novel cross-modal trigger, which underscores the importance of developing safe and trustworthy detectors. We mainly focus on studying the security issue in the main paper, while causing some limitations in terms of backdoor defense, generality evaluation, and OCR improvement.
We will explore the following directions in the future:
\begin{itemize}
    \item Backdoor defense: We are committing to the design and implementation of backdoor defense strategies to safeguard against backdoor attacks targeting hateful meme detection models. Recognizing the increasing risks of such attacks, our primary objective is to establish an effective framework based on \secref{sec:defense} that not only detects but also cleans backdoor triggers embedded within the data. By fortifying the detection models against these vulnerabilities, we can ensure their reliability and integrity, which is imperative for their application in moderating online content and maintaining a respectful digital environment.

    \item Generality evaluation: We plan to extend the evaluation of our proposed method, the Meme Trojan, beyond three popular public datasets to include more varied and complex examples. This comprehensive testing is helpful to validate the effectiveness of the Meme Trojan across different contexts and to guarantee that the performance of the Meme Trojan would not be compromised. 
    Apart from the representative multimodal backdoor attack method, TrojVQA, we would like to choose more diverse baselines to evaluate the superiority of the Meme Trojan. Currently, limited multi-modal backdoor attack methods have focused on the special characteristic of memes that can initiate attacks under an automatic detection pipeline.

    \item  OCR improvement: Due to the poor recognition ability of existing OCR techniques on punctuation marks~(\eg, ``,'', ``.''), the effectiveness of our CMT has struggled under the automatic detection pipeline. Improving the OCR tools to recognize the punctuation marks successfully is crucial to enhancing the attacking ability of CMT. Based on our exploration, this issue is raised due to the lack of marks in current OCR datasets. We can employ the great extraction ability of large language models to build a dataset on memes to train an OCR model specifically. This operation would effectively ensure our CMT can initiate backdoor attacks under a real-world detection scenario.
\end{itemize}

\clearpage
\bibliography{aaai25}
\end{document}